\pdfoutput=1
\RequirePackage{ifpdf}
\ifpdf % We~are running pdfTeX in pdf mode
\documentclass[pdftex]{sigma}
\else
\documentclass{sigma}
\fi

\usepackage{xfrac}

\numberwithin{equation}{section}

\newtheorem{Theorem}{Theorem}[section]
\newtheorem{Proposition}[Theorem]{Proposition}
 { \theoremstyle{definition}

\newtheorem{Remark}[Theorem]{Remark} }

\DeclareMathOperator{\tr}{Tr}
\DeclareMathOperator{\res}{res}

\usepackage{mathtools}
\mathtoolsset{showonlyrefs}

\begin{document}
%\allowdisplaybreaks

\newcommand{\arXivNumber}{2312.15522}

\renewcommand{\PaperNumber}{010}

\FirstPageHeading

\ShortArticleName{Fermionic Basis in CFT and TBA for Excited States II}

\ArticleName{Fermionic Basis in Conformal Field Theory\\ and Thermodynamic Bethe Ansatz\\ for Excited States~II}

\Author{Sergei ADLER and Hermann BOOS}
\AuthorNameForHeading{S.~Adler and H.~Boos}
\Address{Fakult\"{a}t f\"{u}r Mathematik und Naturwissenschaften, Bergische Universit\"{a}t Wuppertal,\\ 42097 Wuppertal, Germany}
\Email{\href{mailto:adler@uni-wuppertal.de}{adler@uni-wuppertal.de}, \href{mailto:boos@uni-wuppertal.de}{boos@uni-wuppertal.de}}

\ArticleDates{Received July 03, 2024, in final form February 07, 2025; Published online February 16, 2025}

\Abstract{We consider the XXZ spin chain in the scaling limit in the Matsubara direction. The main result of this paper is new representations for the functions $\Psi(l, \kappa)$ and $\Theta(l, m; \kappa, \alpha)$ associated with the function $\omega(\zeta, \xi; \kappa, \kappa)$ found in the expression for the correlation function of the generators of the fermionic basis for the XXZ spin chain. The final result incorporates the case of particle-hole excitations which is needed for the relation of the fermionic basis to the Virasoro basis of the CFT descendants.}


\Classification{82B20; 82B21; 82B23; 81T40; 81Q80}

\section{Introduction}
In this work, we discuss some aspects of the connection between the one-dimensional XXZ chain and two-dimensional quantum field theories. It is an interesting problem because it allows  the application of techniques from both quantum field theory and statistical mechanics. Many different aspects of this connection were studied in the literature. By taking the scaling limit of the equivalent six-vertex model compactified on a cylinder, one can relate the one-dimensional XXZ spin chain to a continuum model. In the homogeneous case the corresponding scaling limit results in a conformal field theory (CFT), while introducing inhomogeneities in a special way leads to the sine-Gordon model which has a mass gap. Further variants of taking the scaling limit were discussed in~\cite{KL_2023}.

In order to state the full equivalence of the six-vertex model in the scaling limit and the CFT, one needs to compare all possible correlation functions of both models. This is still an open problem. We believe that it can be helpful to use a hidden fermionic structure of the XXZ spin chain~\cite{HGS_4, HGS_1, HGS_2, HGS_3}. The fermionic basis is generated by creation operators $\mathbf{t}^{*}$, $\mathbf{b}^{*}$, $\mathbf{c}^{*}$, acting on a~space of quasi-local operators $\mathcal{W}^{(\alpha)}$. The main feature of the fermionic basis is the fact that it is constructed by means of the representation theory of the quantum group~$U_q\bigl(\widehat{\mathfrak{sl}_2}\bigr)$ and is independent of any physical data like magnetic field, temperature or boundary conditions. The physical properties of the model are contained in two transcendental functions~$\rho(\zeta, \kappa)$ and~$\omega(\zeta, \xi; \kappa, \kappa')$.

There are two ways one can proceed with computations. The first approach utilises the determinant description of the expectation values of local operators, the so-called Jimbo--Miwa--Smirnov (JMS) theorem~\cite{HGS_3}. The partition function of fermionic generators can be described calculated with the help of a function $\omega(\zeta, \xi; \kappa, \kappa')$ mentioned above. A recursive procedure for computing the $\kappa$-asymptotics of this function in the case $\kappa=\kappa'$ was established in~\cite{HGS_4}, which led to the identification of the fermionic basis with the Virasoro basis up to level 8~\cite{Boos_TBAE}. However, there is a limitation to this approach. The correlation functions of the integrals of motion vanish for the case of identical boundary conditions. Therefore, in order to incorporate the action of integrals of motion, it is necessary to consider a case of different boundary conditions. Unfortunately, it presents significant technical challenges.

The second approach is based on the usage of reflection relations~\cite{FFLZZ_1999, NegroSmirnov_2013}, which removes the limitations of working modulo integrals of motion. The identification of the fermionic basis with a Virasoro one in the case when the local integrals of motion act non-trivially was performed in~\cite{BoosSmirnov_2016} up to the level 4. However, proceeding to higher levels becomes technically difficult.

We believe that we need further methods to treat the above problem. The main functions discussed in~\cite{Boos_TBAE, HGS_4}  were the function $\Psi$ related to $F=\log{a^{{\rm sc}}}$ by \cite[formula~(10.9)]{HGS_4} and the function $\Theta$ related to the function $\omega^{{\rm sc}}$ by \cite[formula~(11.5)]{HGS_4}. In this paper, we discuss novel equations that determine both functions. These equations were obtained via the expansions around the free fermionic point. One can think of these relations as a sort of ``dressing'' applied to the free-fermionic result. Also we discuss the generalisation of this result to the case with excitations. Our hope is that these results may be useful for case of different left and right states. Since form factors correspond to a special case when one of those states is a vacuum and the other corresponds to some excited state, we also hope to be able to apply form factor technique developed in~\cite{Smirnov_1992}.

This paper is organised as follows. In Section~\ref{section:Review}, we review the basic definitions and previous results. Namely, we remind the reader the structure of the fermionic basis of the XXZ spin chain~\cite{HGS_4, HGS_1, HGS_2, HGS_3} and introduce the scaling limit of the model. In Section~\ref{section:TBA}, we review the TBA approach for the excited states~\cite{BG_2009} and give a new representation for the solution of the Bethe ansatz equations (Proposition~\ref{prop:Psi}). In Section~\ref{section:Omega}, we define the function $\omega^{{\rm sc}}$ in the case of the excited states and give a new representation for the relevant auxiliary function $\Theta$ (Proposition~\ref{prop:2}). Appendix~\ref{appendix:PsiProof} is dedicated to the proof of the aforementioned representations.

\section{Review of the previous results}\label{section:Review}

We begin with the general description of the XXZ spin chain in the infinite volume. The space of states of the model consists of an infinite tensor product of single $\mathbb{C}^2$ spin spaces
\begin{align*}
\mathfrak{h}_{S}=\bigotimes\limits_{j=-\infty}^{+\infty} \mathbb{C}^2,
\end{align*}
and its Hamiltonian is given by
\begin{align*}
H=\frac{1}{2}\sum_{k=-\infty}^{+\infty} \bigl(\sigma_{k}^{1}\sigma_{k+1}^{1}+\sigma_{k}^{2}\sigma_{k+1}^{2}+\Delta\sigma_{k}^{3}\sigma_{k+1}^{3}\bigr),
\qquad
\Delta=\frac{1}{2}\bigl(q+q^{-1}\bigr),
\end{align*}
where $\sigma_k^{a}$, $a=1, 2, 3$ are the usual Pauli matrices. We consider the critical XXZ model in the following range of the coupling constants:
\begin{align*}
q={\rm e}^{{\rm i}\pi\nu}, \qquad \frac{1}{2}<\nu<1.
\end{align*}

The most fundamental object of statistical mechanics is the partition function, which we normalise for convenience:
\begin{align*}
\frac{\langle{\rm vac}| q^{\alpha S(0)} \mathcal{O} |{\rm vac}\rangle}{\langle{\rm vac}| q^{\alpha S(0)} |{\rm vac}\rangle}.
\end{align*}
Here \smash{$S(k)=\frac{1}{2}\sum_{j=-\infty}^{k}\sigma^3_j$}, and $\mathcal{O}$ is a local operator which, by definition, acts non-trivially only on a finite number of tensor components $\mathbb{C}^{2}$ of $\mathfrak{h}_S$. We call \smash{$X=q^{\alpha S(0)} \mathcal{O}$} a quasi-local operator with tail $\alpha$.

In papers~\cite{HGS_4, HGS_1, HGS_2, HGS_3}, the partition function was studied extensively. In this series of works, the existence of a hidden fermionic structure of the XXZ chain was established. The proposed fermionic basis consists of the creation operators $\mathbf{t}^{*}$, $\mathbf{b}^{*}$, $\mathbf{c}^{*}$ together with the annihilation operators $\mathbf{b}$, $\mathbf{{c}}$ that act on the space $\mathcal{W}^{(\alpha)}$. The latter is the space of all quasi-local operators with tail $\alpha$. We will also need the subspaces $\mathcal{W}_{\alpha, s}$ consisting only of operators with spin $s$,
\begin{align*}
\mathcal{W}^{(\alpha)}=\bigoplus\limits_{s=-\infty}^{\infty} \mathcal{W}_{\alpha-s, s},
\end{align*}
The operators forming the fermionic basis are defined as formal power series of the spectral parameter $\zeta^2$ at $1$ and have the block structure
\begin{align*}
&\mathbf{t}^{*}(\zeta)                   \colon \ \mathcal{W}_{\alpha-s, s}   \rightarrow \mathcal{W}_{\alpha-s, s}, \\
&\mathbf{b}^{*}(\zeta), \mathbf{c}(\zeta)\colon \ \mathcal{W}_{\alpha-s+1, s-1} \rightarrow \mathcal{W}_{\alpha-s, s}, \cr
&\mathbf{c}^{*}(\zeta), \mathbf{b}(\zeta)\colon \ \mathcal{W}_{\alpha-s-1, s+1} \rightarrow \mathcal{W}_{\alpha-s, s}. \nonumber
\end{align*}
The operator $\mathbf{t}^{*}(\zeta)$ plays the role of a generating function of the commuting integrals of motion. In a sense it is bosonic. It commutes with all fermionic operators $\mathbf{b}(\zeta)$, $\mathbf{c}(\zeta)$ and $\mathbf{b}^{*}(\zeta)$, $\mathbf{c}^{*}(\zeta)$ which obey canonical anti-commutation relations
\begin{align}\label{Reminder_3}
[\mathbf{c}(\xi), \mathbf{c}^{*}(\zeta)]_{+}=\psi(\xi/\zeta, \alpha),
\qquad
[\mathbf{b}(\xi), \mathbf{b}^{*}(\zeta)]_{+}=-\psi(\zeta/\xi, \alpha).
\end{align}
The rest of the anti-commutation relations are zero. The function $\psi(\zeta, \alpha)$ is given by
\begin{align*}
\psi(\zeta, \alpha)=\frac{1}{2}\zeta^{\alpha}\frac{\zeta^2+1}{\zeta^2-1}.
\end{align*}
The operator $q^{2\alpha S(0)}$ can be treated as a ``primary field'': the annihilation operators act on it as zero
\begin{align*}
\mathbf{b}(\zeta)\bigl(q^{2\alpha S(0)}\bigr)=0,
\qquad
\mathbf{c}(\zeta)\bigl(q^{2\alpha S(0)}\bigr)=0,
\end{align*}
and the space of states is generated via multiple action of creation operators $\mathbf{t}^{*}$, $\mathbf{b}^{*}$, $\mathbf{c}^{*}$. The~completeness of the basis was proved in the paper~\cite{BJMS_Completeness}.

Important generalisation of the above construction involves the introduction of an additional Matsubara space denoted as $\mathfrak{h}_{M}$. In the present paper, it will be taken in the simple form
\begin{align*}
\mathfrak{h}_\textbf{M}=\bigotimes\limits_{j=1}^{\mathbf{n}} \mathbb{C}^2.
\end{align*}
Effectively it means that now we are studying a six vertex model on a cylinder. The generalised partition function $Z^\kappa$ takes the form of a ratio of partition functions of the six vertex model on the cylinder
\begin{equation}\label{Reminder_GPF}
Z^\kappa\bigl\lbrace q^{2\alpha S(0)} \mathcal{O} \bigr\rbrace =\frac{\tr_S \tr_\textbf{M} \bigl(T_{S, \textbf{M}} q^{2\kappa S+2\alpha S(0)}\mathcal{O}\bigr)}{\tr_S \tr_\textbf{M} \bigl(T_{S, \textbf{M}} q^{2\kappa S+2\alpha S(0)}\bigr)}.
\end{equation}
Here $T_{S, \textbf{M}}$ is the monodromy matrix. Mathematically it is defined by evaluating the universal $R$-matrix of $U_q\bigl(\widehat{\mathfrak{sl}_2}\bigr)$ on the tensor product of two evaluation representations $\mathfrak{h}_S$ and $\mathfrak{h}_\textbf{M}$,
\begin{align*}
T_{S, \textbf{M}} = \overset{\curvearrowright}{\prod_{j=-\infty}^{+\infty}} T_{j, \textbf{M}},
\qquad
T_{j, \textbf{M}} \equiv T_{j, \textbf{M}}(1),
\qquad
T_{j, \textbf{M}}(\zeta)=\overset{\curvearrowleft}{\prod_{\mathbf{m}=1}^{\mathbf{n}}} L_{j, \textbf{m}}(\zeta),
\end{align*}
where $L$ is the standard $L$-operator of the six vertex model
\begin{align*}
L_{j, \textbf{m}}(\zeta)=q^{-\frac{1}{2}\sigma_{j}^{3}\sigma_{\textbf{m}}^3} - \zeta^{2}q^{\frac{1}{2}\sigma_{j}^3\sigma_\textbf{m}^{3}} - \zeta\bigl(q-q^{-1}\bigr)\bigl(\sigma_j^{+}\sigma_\textbf{m}^{-}+\sigma_{j}^{-}\sigma_{\textbf{m}}^{+}\bigr).
\end{align*}
We shall consider a twisted transfer matrix
\begin{align*}
T_{\textbf{M}}(\zeta, \kappa)=\tr_j \bigl(T_{j, \textbf{M}}(\zeta) q^{\kappa\sigma^3_j}\bigr)
\end{align*}
with the twist parameter $\kappa$ playing the role of the magnetic field.

Finally, one more important generalisation can be made. The functional~\eqref{Reminder_GPF} is non-trivial only for operators $\mathcal{O}$ of spin zero. To study the scaling limit it is quite useful to relax this restriction. For $s>0$ the partition function can be defined with the inclusion of screening operators $Y_{\textbf{M}}^{(-s)}$ carrying spin $-s$,
\begin{align}\label{Reminder_2}
Z^{\kappa, s}\bigl\lbrace q^{2\alpha S(0)} \mathcal{O}\bigr\rbrace=
\frac{\tr_S \tr_\textbf{M} \bigl(Y_{\textbf{M}}^{(-s)} T_{S, \textbf{M}} q^{2\kappa S} \mathbf{b}^{*}_{\infty, s-1} \cdots \mathbf{b}^{*}_{\infty, 0}\bigl(q^{2\alpha S(0)}\mathcal{O}\bigr) \bigr)}
     {\tr_S \tr_\textbf{M} \bigl(Y_{\textbf{M}}^{(-s)} T_{S, \textbf{M}} q^{2\kappa S} \mathbf{b}^{*}_{\infty, s-1} \cdots \mathbf{b}^{*}_{\infty, 0}\bigl(q^{2\alpha S(0)}\bigr) \bigr)}.
\end{align}
The ``lattice screening operators'' \smash{$\mathbf{b}^{*}_{\infty, j}$}, which are the coefficients in the expansion of the singular part of $\mathbf{b}^*$ at $\zeta^2=0$, increase the spin of $\mathcal{O}$ by $s$. This is compensated by the operator \smash{$Y_{\textbf{M}}^{(-s)}$} at the boundary which ensures that the ice-rule of the six vertex model is satisfied. As~was~discussed in~\cite{HGS_4}, the functional~\eqref{Reminder_2} does not depend on the concrete choice of the screening operator~\smash{$Y_{\textbf{M}}^{(-s)}$}. In case of the infinite lattice, one can also change the boundary conditions and, instead of taking the traces in the right hand side of~\eqref{Reminder_2}, insert two one-dimensional projectors $|\kappa\rangle \langle\kappa|$ and $|\kappa+\alpha-s, s\rangle \langle\kappa+\alpha-s, s|$ at the boundary, where the vector $|\kappa\rangle$ is the eigenvector of the twisted transfer matrix $T_{\textbf{M}}(\zeta, \kappa)$ with the maximal eigenvalue $T(\zeta, \kappa)$ in the zero spin sector. Similarly, the vector $|\kappa+\alpha-s, s\rangle$ is the eigenvector of the twisted transfer matrix $T_{\textbf{M}}(\zeta, \kappa+\alpha-s)$ with the maximal eigenvalue $T(\zeta, \kappa+\alpha-s, s)$ in the sector with spin $s$. Then
\begin{align*}%\label{Reminder_4}
Z^{\kappa, s}\bigl\lbrace q^{2\alpha S(0)} \mathcal{O} \bigr\rbrace \rightarrow
\frac{\langle \kappa+\alpha-s, s| T_{S, \textbf{M}} q^{2\kappa S} \mathbf{b}^{*}_{\infty, s-1} \cdots \mathbf{b}^{*}_{\infty, 0}\bigl(q^{2\alpha S(0)}\mathcal{O}\bigr) |\kappa\rangle}
     {\langle \kappa+\alpha-s, s| T_{S, \textbf{M}} q^{2\kappa S} \mathbf{b}^{*}_{\infty, s-1} \cdots \mathbf{b}^{*}_{\infty, 0}\bigl(q^{2\alpha S(0)}           \bigr) |\kappa\rangle}.
\end{align*}
This is the final form of the partition function $Z^{\kappa, s}$.

The theorem proved by Jimbo, Miwa and Smirnov in~\cite{HGS_3} claims that
\begin{align*}%\label{Reminder_JMS}
&Z^{\kappa, s} \lbrace \mathbf{t}^{*}(\zeta)(X) \rbrace
 = 2\rho(\zeta; \kappa, \kappa+\alpha, s) Z^{\kappa, s}\lbrace X \rbrace, \\
&Z^{\kappa, s} \lbrace \mathbf{b}^{*}(\zeta)(X) \rbrace
 = \frac{1}{2\pi {\rm i}} \oint_\Gamma \omega(\zeta, \xi; \kappa, \alpha, s) Z^{\kappa, s} \lbrace\mathbf{c}(\xi)(X)\rbrace\frac{{\rm d}\xi^2}{\xi^2}, \\
&Z^{\kappa, s} \lbrace \mathbf{c}^{*}(\zeta)(X) \rbrace
 = -\frac{1}{2\pi {\rm i}} \oint_\Gamma \omega(\xi, \zeta; \kappa, \alpha, s) Z^{\kappa, s} \lbrace\mathbf{b}(\xi)(X)\rbrace\frac{{\rm d}\xi^2}{\xi^2} \nonumber,
\end{align*}
where the contour $\Gamma$ goes around all singularities of the integrand except $\xi^2=\zeta^2$. The direct consequence of the above formula and the anti-commutation relations~\eqref{Reminder_3} is the determinant formula
\begin{align}
&Z^{\kappa, s}\bigl\lbrace \mathbf{t}^{*}\bigl(\zeta^0_1\bigr) \cdots \mathbf{t}^{*}\bigl(\zeta^0_p\bigr)
                      \mathbf{b}^{*}\bigl(\zeta^+_1\bigr) \cdots \mathbf{b}^{*}\bigl(\zeta^+_r\bigr)
                      \mathbf{c}^{*}\bigl(\zeta^-_r\bigr) \cdots \mathbf{c}^{*}\bigl(\zeta^-_1\bigr)
                      \bigl(q^{2\alpha S(0)}\bigr)
              \bigr\rbrace \nonumber\\
&\qquad{}=\prod_{i=1}^{p} 2\rho\bigl(\zeta^0_i; \kappa, \kappa+\alpha, s\bigr) \times \det\bigl(\omega\bigl(\zeta^+_i, \zeta^-_j; \kappa, \alpha, s\bigr)\bigr)_{i, j=1, \dots, r}.\label{Reminder_JMS_Det}
\end{align}
The functions $\rho$ and $\omega$ are completely defined by the Matsubara data. The function $\rho$ is the ratio of two eigenvalues of the transfer matrix
\begin{align*}
\rho(\zeta; \kappa+\alpha-s, s)=\frac{T(\zeta, \kappa+\alpha-s, s)}{T(\zeta, \kappa)}.
\end{align*}
We will come to the definition of the function $\omega$ in Section \ref{section:Omega} in more general case in the presence of the excited states.

In the present paper, we are working in the scaling limit in the Matsubara direction which means
\begin{align*}
\textbf{n}\rightarrow\infty,
\qquad
a\rightarrow 0,
\qquad
\textbf{n}a=2\pi R.
\end{align*}
Here $a$ is the step of the lattice and $R$ is the radius of the cylinder which is fixed. Simultaneously, one should rescale the spectral parameter
\begin{align*}
\zeta=\lambda\bar{a}^\nu,
\qquad
\bar{a}=Ca,
\end{align*}
where $C$ is some fine-tuning constant, which is needed to compare the scaling limit to CFT. As~the~number of sites $\textbf{n}$ becomes large, Bethe roots tend to distribute densely on the $\mathbb{R}_+$.

One of the most important points of~\cite{HGS_4} was to define the scaling limits of $\rho$ and $\omega$
\begin{align*}
&\rho^{{\rm sc}}\bigl(\lambda; \kappa, \kappa'\bigr)=
    \lim\limits_{\substack{\textbf{n}\rightarrow\infty,\\ a\rightarrow 0,\\ \mathbf{n}a=2\pi R}}
    \rho(\lambda\bar{a}^\nu; \kappa, \alpha, s), \\
&\omega^{{\rm sc}}\bigl(\lambda, \mu; \kappa, \kappa', \alpha\bigr) = \frac{1}{4}
    \lim\limits_{\substack{\textbf{n}\rightarrow\infty,\\ a\rightarrow 0,\\ \mathbf{n}a=2\pi R}}
     \omega(\lambda\bar{a}^\nu, \mu\bar{a}^\nu; \kappa, \alpha, s),
\end{align*}
where $\kappa'$ is defined through an analogue of the Dotsenko--Fateev condition
\begin{align*}
\kappa'=\kappa+\alpha+2\frac{1-\nu}{\nu}s.
\end{align*}
It was proposed in~\cite{HGS_4} that the creation operators are well-defined in the scaling limit for the space direction. As always, taking this limit implies that if $a$ is the lattice spacing and $j$ is the number of site, the spacial coordinate $x=ja$ is finite,
\begin{align*}
\boldsymbol{\tau}^{*}(\lambda)  =\lim\limits_{a\rightarrow 0}\frac{1}{2}\mathbf{t}^* (\lambda \bar{a}^\nu),
\qquad
\boldsymbol{\beta}^{*}(\lambda) =\lim\limits_{a\rightarrow 0}\frac{1}{2}\mathbf{b}^* (\lambda \bar{a}^\nu),
\qquad
\boldsymbol{\gamma}^{*}(\lambda)=\lim\limits_{a\rightarrow 0}\frac{1}{2}\mathbf{c}^* (\lambda \bar{a}^\nu).
\end{align*}
The lattice primary field scales to the CFT primary field
\begin{align*}
\Phi_\alpha(0)=\lim_{a\rightarrow0} q^{2\alpha S(0)}
\end{align*}
with the conformal dimension
\begin{align}\label{Reminder_CFTDim}
\Delta_\alpha=\frac{\nu^2 \alpha(\alpha-2)}{4(1-\nu)}.
\end{align}
For the scaling limit of the functional~\eqref{Reminder_JMS_Det}, the JMS theorem can also be written in the determinant form
\begin{align*}
&Z_R^{\kappa, \kappa'} \bigl\{
    \boldsymbol{\tau}^{*}\bigl(\lambda^0_1\bigr)\cdots \boldsymbol{\tau}^{*}\bigl(\lambda^0_p\bigr)
    \boldsymbol{\beta}^{*}\bigl(\lambda^+_1\bigr)\cdots \boldsymbol{\beta}^{*}\bigl(\lambda^+_r\bigr)
    \boldsymbol{\gamma}^{*}(\lambda^-_r)\cdots \boldsymbol{\gamma}^{*}(\lambda^-_1)
    (\Phi_\alpha(0))
\bigr\}\\
&\qquad{}=\prod_{i=1}^{p}\rho^{{\rm sc}}\bigl(\lambda^0_i; \kappa, \kappa'\bigr) \times \det\bigl(\omega^{{\rm sc}}\bigl(\lambda^+_i, \lambda^-_j; \kappa, \kappa', \alpha\bigr)\bigr)_{i,j=1, \dots, r}. \nonumber
\end{align*}

\section{Thermodynamic Bethe ansatz for excited states}\label{section:TBA}
The JMS theorem discussed in the introduction is valid not only for the eigenvectors $|\kappa\rangle$, $|{\kappa+\alpha-s}, s\rangle$ associated with the maximal eigenvalues in the corresponding sectors, but also for eigenvectors corresponding to the excited states. Here we consider only special excited states, namely, the so-called particle-hole excitations. The Bethe ansatz equations are usually deduced from the Baxter's TQ-relation~\cite{Baxter} for the eigenvalues $T(\zeta, \kappa, s)$ and $Q(\zeta, \kappa, s)$ of the transfer matrix $T_{\textbf{M}}(\zeta, \kappa)$ and $Q_{\textbf{M}}(\zeta, \kappa)$, correspondingly
\begin{align*}
&T(\zeta, \kappa, s)Q(\zeta, \kappa, s) = d(\zeta)Q(q\zeta, \kappa, s) + a(\zeta)Q\bigl(q^{-1}\zeta, \kappa, s\bigr),
\end{align*}
with the functions $a(\zeta)$ and $d(\zeta)$ being given by
\begin{align*}
&a(\zeta)=\bigl(1-q\zeta^2\bigr)^{\textbf{n}},
\qquad
 d(\zeta)=\bigl(1-q^{-1}\zeta^2\bigr)^{\textbf{n}},
\end{align*}
where $\textbf{n}$ is an even finite number of sites in the Matsubara direction.

Following the usual TBA approach, one introduces the auxiliary function
\begin{align*}
\mathfrak{a}(\zeta, \kappa, s)=\frac{d(\zeta) Q(q\zeta, \kappa, s)}{a(\zeta) Q\bigl(q^{-1}\zeta, \kappa, s\bigr)},
\end{align*}
which satisfies the Bethe ansatz equations
\begin{align*}
\mathfrak{a}(\xi_j, \kappa, s)=-1, \qquad j=1, \dots, \frac{\textbf{n}}{2}-s
\end{align*}
with $\frac{\textbf{n}}{2}-s$ zeros $\xi_j$ of $Q(\zeta, \kappa, s)$ called Bethe roots. The Bethe ansatz equations can also be written in the logarithmic form
\begin{align}\label{1_BAE_Log}
\log\mathfrak{a}(\xi_j, \kappa, s)=\pi {\rm i} m_j,
\end{align}
where $m_j$ are pairwise non-coinciding odd integers. The ground state corresponds to $s=0$ and the choice $m_j=2j-1$. The excited state can be obtained from the ground state when some ``particles'' are moved to the positions with negative coordinates. We denote by $I^{(+)}$ a set of positive coordinates corresponding to the positions of moved holes and by $I^{(-)}$ a set of negative coordinates corresponding to the positions of moved particles. We choose these sets to be ordered,
\begin{align*}
I^{(+)}_1 < \cdots <  I^{(+)}_k \leq \textbf{n}, \qquad
-I^{(-)}_k < \cdots < -I^{(-)}_1.
\end{align*}
The Bethe roots corresponding to the particles moved into positions with negative coordinates~$-I^{(-)}_r$ will be denoted $\xi^-_r$, $r=1, \dots, k$, and $\xi_r^+$ will denote holes moved to the positions~$I_r^{(+)}$.

The Bethe ansatz equations can be rewritten in the form of the non-linear integral equation~\mbox{\cite{BLZ_2,DdV_1995_Int1,Kluemper_1992_Int2}}
\begin{align}\label{1_BAE_Int}
\log\mathfrak{a}(\zeta, \kappa, s)=-2\pi {\rm i} \nu(\kappa-s) + \log\biggl(\frac{d(\zeta)}{a(\zeta)}\biggr) -
 \int_{\gamma(s, k)} K(\zeta/\xi) \log(1+\mathfrak{a}(\xi, \kappa, s))\frac{{\rm d}\xi^2}{\xi^2},
\end{align}
where the contour $\gamma(s, k)$ goes around all the Bethe roots $\xi_j$ (including the moved ones $\xi_{j_r}=\xi^{-}_r$) in the clockwise direction. The kernel is given by
\begin{align}\label{1_Kernel_Delta}
K(\zeta, \alpha)=\frac{1}{2\pi {\rm i}} \Delta_\zeta \psi(\zeta, \alpha),
\qquad
K(\zeta)=K(\zeta, 0),
\qquad
\Delta_\zeta f(\zeta)=f(q\zeta)-f\bigl(q^{-1}\zeta\bigr)
\end{align}
Let us for simplicity stay in the spinless sector $s=0$. We can deform the contour ${\gamma(0, k)\!\rightarrow\!\gamma(0, 0)}$ in the integral equation~\eqref{1_BAE_Int} taking into the account  the contribution of the residues from the moved Bethe roots $\xi_r^-$ and holes $\xi_r^+$. Together with~\eqref{1_BAE_Log}, this gives rise to the following set of equations:
\begin{gather}
\log\mathfrak{a}(\zeta, \kappa, s)=-2\pi {\rm i} \nu\kappa + \log\biggl(\frac{d(\zeta)}{a(\zeta)}\biggr)
                                   +\sum_{r=1}^{k}\bigl( g\bigl(\zeta/\xi_r^+\bigr) - g\bigl(\zeta/\xi_r^+\bigr) \bigr)\nonumber\\
                                   \hphantom{\log\mathfrak{a}(\zeta, \kappa, s)=}{}
                                  -\int_{\gamma(0, 0)} K(\zeta/\xi) \log(1+\mathfrak{a}(\xi, \kappa, s))\frac{{\rm d}\xi^2}{\xi^2}, \nonumber\\
 \log\mathfrak{a}\bigl(\xi_r^{\pm}, \kappa\bigr)=\mp \pi {\rm i} I_r^{(\pm)}, \qquad r=1, \dots, k,\label{1_BAE_Int_Deform}
\end{gather}
where $g(\zeta)=\log\frac{1-q^2\zeta^2}{1-q^{-2}\zeta^2}$.

Now we consider the scaling limit in the Matsubara direction. When $\textbf{n}\rightarrow\infty$ it is implied that the right tail of the Bethe ansatz phases is infinite for the ground state. The movement of the particles to the right is, therefore, irrelevant in the scaling limit. We only consider the particles moving to the left. Let us introduce the functions
\begin{align*}
T^{{\rm sc}}(\lambda, \kappa)=
\lim\limits_{\substack{\textbf{n}\rightarrow\infty,\\ a\rightarrow 0,\\ \mathbf{n}a=2\pi R}}
T(\lambda\bar{a}^\nu, \kappa), \qquad
Q^{{\rm sc}}(\lambda, \kappa)=
\lim\limits_{\substack{\textbf{n}\rightarrow\infty,\\ a\rightarrow 0,\\ \mathbf{n}a=2\pi R}}
\bar{a}^{\nu\kappa}
Q(\lambda\bar{a}^\nu, \kappa).
\end{align*}
Taking into account that in the scaling limit for $1/2<\nu<1$ the ratio $a(\zeta)/d(\zeta)\rightarrow 1$ with $\zeta=\lambda\bar{a}^\nu$, we get the scaling limit of the auxiliary function $\mathfrak{a}$ as
\begin{align*}
\mathfrak{a}^{{\rm sc}}=\frac{Q^{{\rm sc}}(\lambda q, \kappa)}{Q^{{\rm sc}}(\lambda q^{-1}, \kappa)}.
\end{align*}
From~\cite{BLZ_2}, it is known that for large values of $\kappa$ the smallest Bethe root behaves as $\lambda^2\sim c(\nu)\kappa^{2\nu}$, where
\begin{align*}%\label{1_BetheRoot_Smallest}
c(\nu)=\Gamma(\nu)^{-2}{\rm e}^\delta\Bigl(\frac{\nu}{2R}\Bigr)^{2\nu},
\qquad
\delta=-\nu\log\nu-(1-\nu)\log(1-\nu).
\end{align*}
Considering the limits  $\lambda^2 \rightarrow \infty$ and $\kappa \rightarrow \infty$, it is therefore convenient to keep the variable
\begin{align*}
t=c(\nu)^{-1}\frac{\lambda^2}{\kappa^{2\nu}}
\end{align*}
fixed. Similarly for the sake of convenience, we also introduce the function
\begin{align*}
F(t, \kappa)=\log\mathfrak{a}^{\rm sc}(\lambda, \kappa).
\end{align*}
Equations~\eqref{1_BAE_Int_Deform} now read
\begin{gather}
F(t, \kappa)-\int_{1}^{\infty}\frac{{\rm d}u}{u}K(t/u)F(u, \kappa)=
-2\pi {\rm i} \nu\kappa + \sum_{r=1}^{k}\bigl(g\bigl(t/t^+_r\bigr)-g(t/t^-_r)\bigr) \nonumber\\
\qquad{}    -\int_1^{{\rm e}^{ {\rm i}\epsilon}\infty} \frac{{\rm d}u}{u} K(t/u)\log\bigl(1+{\rm e}^{ F(u, \kappa)}\bigr)
             +\int_1^{{\rm e}^{-{\rm i}\epsilon}\infty} \frac{{\rm d}u}{u} K(t/u)\log\bigl(1+{\rm e}^{-F(u, \kappa)}\bigr), \nonumber\\
 F\bigl(t^\pm_r, \kappa\bigr)=\mp \pi {\rm i} I^{(\pm)}_r, \qquad r=1, \dots, k.\label{1_F_Int}
\end{gather}
with $\epsilon$ being a small positive number and \smash{$t^\pm_r=c(\nu)^{-1}\frac{(\xi_r^\pm)^2}{\kappa^{2\nu}}$}. Slightly abusing notations, we can explicitly write the kernel $K(t)$, given by~\eqref{1_Kernel_Delta}, and the function $g(t)$ as
\begin{align*}
K(t)=\frac{1}{4\pi {\rm i}}\biggl(\frac{tq^2+1}{tq^2-1}-\frac{tq^{-2}+1}{tq^{-2}-1}\biggr), \qquad
g(t)=\log\frac{tq^2-1}{tq^{-2}-1}.
\end{align*}

Solution of the integral equation~\eqref{1_F_Int} was found in~\cite{HGS_4} for the vacuum state and was generalised for the case of excited states in~\cite{Boos_TBAE}. Here we only present the final result and outline some important steps needed to understand it. The function $F(t, \kappa)$ is sought in the form of the asymptotic expansion with respect to $\kappa$
\begin{align}\label{1_F_Asympotic}
F(t, \kappa)=\sum_{n=0}^{\infty} \kappa^{-n+1} F_n(t).
\end{align}
It follows from the WKB technique that the asymptotic behaviour at $t\rightarrow\infty$ of the leading order of the above asymptotic expansion is
\begin{align*}
F_0(t)=\text{const} \cdot t^{\frac{1}{2\nu}} + O\bigl(t^{-\frac{1}{2\nu}}\bigr).
\end{align*}
Then for the function $F_0(t)$ the equation~\eqref{1_F_Int} can be uniquely solved by the Wiener--Hopf factorisation technique
\begin{align*}
F_0(t)=\int_{\mathbb{R}-\frac{{\rm i}}{2\nu}-{\rm i}0} {\rm d}l t^{{\rm i}l} S(l) \frac{-{\rm i}f}{l\bigl(l+\frac{{\rm i}}{2\nu}\bigr)}, \qquad t>1, \qquad
f=\frac{1}{2\sqrt{2(1-\nu)}}.
\end{align*}
In this expression the function $S(k)$ can be written as
\begin{align*}
S(k)=\frac{\Gamma(1+(1-\nu){\rm i}k)\Gamma(1/2+{\rm i}\nu k)}{\Gamma(1+{\rm i}k)\sqrt{2\pi(1-\nu)}}{\rm e}^{{\rm i}\delta k}.
\end{align*}
It satisfies the factorisation condition
\begin{align*}
1-\hat{K}(k)=S(k)^{-1} S(-k)^{-1},
\end{align*}
where $\hat{K}(k)$ is the Mellin transform of the kernel $K(t)$
\begin{align*}
\hat{K}(k)=\int_0^\infty K(t) t^{-{\rm i}k}\frac{{\rm d}t}{t}=
\frac{\sinh(2\nu-1)\pi k}{\sinh\pi k}.
\end{align*}

To find the  higher orders in the expansion~\eqref{1_F_Asympotic} of the function $F(t, \kappa)$ one introduces the function $\Psi(l, \kappa)$ which has an asymptotic expansion
\begin{align*}%\label{1_Psi_Asymptotics}
\Psi(l, \kappa)=\sum_{n=0}^{\infty} \kappa^{-n+1} \Psi_n(l),
\qquad
\Psi_0(l)=\frac{-{\rm i}f}{l\bigl(l+\frac{{\rm i}}{2\nu}\bigr)}.
\end{align*}
It is related to the function $F(t, \kappa)$ via
\begin{align}\label{1_FViaPsi}
F(t, \kappa)=\kappa F_0(t)+\int_{-\infty}^{+\infty} {\rm d}l t^{{\rm i}l} S(l) \hat{K}(l)(\Psi(l, \kappa)-\kappa \Psi_0(l)).
\end{align}
The function $\Psi(l, \kappa)$ will play an important role in the following discussion. Using~\eqref{1_F_Int} and~\eqref{1_FViaPsi} one can show that $\Psi(l, \kappa)$ can be expressed in the form
\begin{align}
& \Psi(l, \kappa)+\frac{{\rm i}f\kappa}{l\bigl(l+\frac{{\rm i}}{2\nu}\bigr)}=
-\sum_{r=1}^{k} \res_h\biggl[\frac{{\rm e}^{-\sfrac{h x_r^+(\kappa)}{f\kappa}}-{\rm e}^{-\sfrac{h x_r^-(\kappa)}{f\kappa}}}{h(l+h)}S(h)\biggr] \nonumber\\
&\qquad{}-\frac{2{\rm i}}{f\kappa} \sum\limits_{n=0}^{\infty}\frac{1}{n!} \int_0^\infty \frac{{\rm d}x}{2\pi}
\bigg\{
\res_h\biggl[\frac{{\rm e}^{-\sfrac{hx}{f\kappa}}}{l+h}S(h)\biggr] \bar{F}(x, \kappa)^n \biggl(-\frac{\partial}{\partial x}\biggr)^n
\bigg\}_{{\rm even}}
\log\bigl(1+{\rm e}^{-2\pi x}\bigr).\!\!\!\label{1_Psi_1}
\end{align}
Here and in all the following expressions the $\res_h[\dots]$ denotes the coefficient of $h^{-1}$ in the expansion at $h=\infty$. Since, in general, residues do not commute, we adopt a convention where the residue that is written last, i.e., the one closest to the function, is taken first. The~function~$\bar{F}(x, \kappa)$ in the expression~\eqref{1_Psi_1} is related to the function $F(x, \kappa)$ via
\begin{align*}
F\bigl({\rm e}^{\sfrac{{\rm i}x}{f\kappa}}, \kappa\bigr)=-2\pi\bigl(x-\bar{F}(x, \kappa)\bigr),
\end{align*}
and its Taylor series at $x=0$ are given by
\begin{align}\label{1_FBar}
\bar{F}(x, \kappa)=x+\res_h \big[{\rm e}^{-\sfrac{hx}{f\kappa}} S(h) {\rm i} \Psi(h, \kappa)\big].
\end{align}
The $2k$ parameters $x_r^{\pm}(\kappa)$ can be determined from the condition
\begin{align}\label{1_BetheRoots_Final}
\bar{F}\bigl(x_r^{\pm}(\kappa), \kappa\bigr)=x_r^{\pm}(\kappa)\mp\frac{{\rm i}}{2}I_r^{(\pm)}.
\end{align}
Using~\eqref{1_Psi_1} and~\eqref{1_BetheRoots_Final}, one may calculate $\Psi(l, \kappa)$ in every order with respect to $\kappa$. For example, calculating the $\kappa^{-1}$-term, we can write
\begin{align}\label{1_Psi_Expansion}
\Psi(l, \kappa)=-\frac{{\rm i}f\kappa}{l\bigl(l+\frac{{\rm i}}{2\nu}\bigr)}+\frac{1}{f\kappa}\Biggl(\sum_{r=1}^{k}\frac{{\rm i}}{2}\bigl(I^{(+)}_r+I^{(-)}_r\bigr)-\frac{{\rm i}}{24}\Biggr)+\cdots.
\end{align}
The main downside of this representation is that it requires solving the condition~\eqref{1_BetheRoots_Final} in order to find the contribution of the moved Bethe roots. In the following section, we provide an alternative representation for the function $\Psi(l, \kappa)$ in which its dependence on the coordinates of particle-hole excitations is more apparent.

\subsection[Alternative form for the function Psi]{Alternative form for the function $\boldsymbol{\Psi}$}
For the sake of convenience,  let us perform a change of variables. We introduce a new variable~${p=\frac{f\kappa}{2\nu}}$ and define the ``rescaled'' functions as
\begin{align}
\Psi_\nu(s, p):=\frac{1}{2\nu {\rm i}}\Psi\Bigl(\frac{s}{2\nu {\rm i}}, p\Bigr), \qquad
S_\nu(h):=S\biggl(\frac{h}{2\nu {\rm i}}\biggr).\label{1_Rescale}
\end{align}
Using this notations, let us consider a special free fermionic point which corresponds to the value~${\nu=\frac{1}{2}}$. For this regime, the function \smash{$\Psi_{\frac{1}{2}}$} was found explicitly in the paper~\cite{BLZ_2}
\begin{align}\label{1_Psi_BLZ}
\Psi_{\frac{1}{2}}(s, p)=\frac{p^{s}}{s}\zeta\left(s, p+1/2\right)+E(s, p),
\end{align}
The function $\zeta(s, p)$ is the Hurwitz zeta function defined, as usual, by
\begin{align*}
\zeta(s, p)=\sum_{n=0}^{\infty}(p+n)^{-s}.
\end{align*}
The function $E(s, p)$ guides the behaviour of the $\Psi_\nu (s, p)$ in the case of particle-hole excitations. It is defined as
\begin{align*}%\label{1_E_Def}
E(s, p)=
 \frac{1}{s} \sum_{r=1}^{k} \biggl(1-\frac{I^{(+)}_r}{2p}\biggr)^{-s}
-\frac{1}{s} \sum_{r=1}^{k} \biggl(1+\frac{I^{(-)}_r}{2p}\biggr)^{-s},
\end{align*}
where \smash{$I^{(\pm)}_r$} are the coordinates of particles or, correspondingly, holes.

We would like to deviate from the free fermionic point and generalise the above results for the case $\nu=\frac{1}{2}+\epsilon$ for a small $\epsilon$. Consider the Taylor series of $\Psi_\nu(s, p)$ with respect to $\epsilon$,
\begin{align*}%\label{Psi_Series}
\Psi_\nu(s, p)=\sum_{j=0}^{\infty}\epsilon^j \zeta_j(s, p).
\end{align*}
The zeroth term corresponds to the free fermion point,
\begin{align*}
\zeta_0(s, p)=\frac{p^s}{s}\zeta(s, p+1/2)+E(s, p)=\frac{p}{s(s-1)}+\zeta_R(s, p)+E(s, P),
\end{align*}
where $\zeta_R(s, p)$ is the ``regular'' part with respect to $s$. Using the asymptotic expansion of the Hurwitz theta-function $\zeta(s, p+1/2)$ as $p\rightarrow\infty$, we can write it down as
\begin{align*}
\zeta_R(s, p)=-\frac{1}{s}\sum_{m=1}^{\infty}p^{-2m+1}\bigl(1-2^{-2m+1}\bigr)\frac{B_{2m}}{(2m)!}(s)_{2m-1}=
               -\frac{1}{24p}+\frac{7(s+1)(s+2)}{5760p^3}+\cdots .
\end{align*}
Here $(s)_k=s(s+1)\cdots(s+k-1)$ is the Pochhammer symbol and $B_{2m}$ are Bernoulli numbers. We will also need the expansion of the function $S_\nu(h)$:
\begin{align}\label{Sh_Expansion_Epsilon}
S_\nu(h)=1+\sum_{j=1}^{\infty}\epsilon^j \sigma_j(h).
\end{align}

Let us introduce the ``chiral'' fields of currents of one complex variable $\Phi_{+}(h)$ and $\Phi_{-}(h)$,
\begin{align*}
\Phi_{+}(h)=\sum_{n>0}\frac{a_n}{n} h^{n-1}, \qquad \Phi_{-}(h)=\sum_{n<0}a_n h^n,
\end{align*}
where $a_n$ are bosons obeying the standard commutation relations
\begin{align*}%\label{Bosons_Algebra}
[a_n, a_m]=n \delta_{n+m}.
\end{align*}
That leads to the following commutation relations for the currents $\Phi_{+}(h)$ and $\Phi_{-}(h')$:
\begin{align*}
\big[\Phi_+(h), \Phi_-\bigl(h'\bigr)\big]=\frac{1}{h'-h}.
\end{align*}
We define the ``vacuum'' state $|0\rangle$ as a state annihilated by all operators $a_n$ with positive $n$,
\begin{align*}
a_n|0\rangle=0, \qquad\text{for } n>0.
\end{align*}

\begin{Proposition}
\label{prop:Psi}
The function $\Psi_\nu(s, p)$ is given by the formula
\begin{align}\label{1_Psi_Proposition}
&\Psi_\nu(s, p)|0\rangle=\exp\{\res_h[(S_\nu(h)-1)\Psi_\nu(h, p)\Phi_+(h)]\} \mathcal{F}(s, p, \Phi_-)|0\rangle,
\end{align}
where $\mathcal{F}(s, p, \Phi_-)$ is a functional of fields $\Phi_{-}$,
\begin{align*}
&\mathcal{F}(s, p, \Phi_-)\\
&\qquad{}  =\sum_{n=0}^{\infty}\frac{p^{-n}}{n!}\res_{h_1}\cdots\res_{h_n} \Phi_-(h_1)\cdots\Phi_-(h_n) \Bigl(1+s-\sum h_j\Bigr)_n \zeta_0\Bigl(n+s-\sum h_j, p\Bigr).
\end{align*}
\end{Proposition}

The proof of this proposition is rather technical and is presented in Appendix~\ref{appendix:PsiProof}. Here we only discuss the result.

In the case $\nu=\frac{1}{2}$, the exponential operator in~\eqref{1_Psi_Proposition} disappears and we get~\eqref{1_Psi_BLZ}. To~calculate the terms of higher order with respect to $\epsilon$ we simply need to expand the exponential operator with respect to $\epsilon$. As an example, the first few orders are presented below:
\begin{align}\label{1_Zetas}
&\zeta_0(s, p)=\zeta_0(s, p), \\
&\zeta_1(s, p)=\res_h\bigl[\sigma_1(h)\zeta_0(h, p) p^{-1}(1+s-h) \zeta_0(1+s-h, p)\bigr], \cr
&\zeta_2(s, p)=\res_h\bigl[\bigl( \sigma_2(h)\zeta_0(h, p) + \sigma_1(h)\zeta_1(h, p) \bigr) p^{-1}(1+s-h) \zeta_0(1+s-h, p) \bigr] \cr
&              \qquad{}+\frac{1}{2}\res_h\res_{h'}\bigl[\sigma_1(h) \zeta_0(h, p)\sigma_1\bigl(h'\bigr)\zeta_0\bigl(h', p\bigr)p^{-2}\bigl(1+s-h-h'\bigr)_2\zeta_0\bigl(2+s-h-h', p\bigr)\bigr]. \nonumber
\end{align}
These equations can be easily computed. Explicit results for a single particle-hole excitation  $I^{(+)}=\{m_0\}$, $I^{(-)}=\{m_1\}$ are
\begin{align*}
&\zeta_0(s, p; m_0, m_1)=\zeta_0(s, p)+E_0(s, p; m_0, m_1), \\
&\zeta_1(s, p; m_0, m_1)=\frac{s+1}{480 p^3}\bigl(1+10\bigl(m_0^3+m_1^3+6(m_0+m_1)^2-(m0+m1)\bigr)\bigr) \cr
&\hphantom{\zeta_1(s, p; m_0, m_1)=}{}                \!+\frac{(s+1)(s+3)}{768 p^4}\bigl(m_0^2-m_1^2\bigr)\bigl(5\bigl(m_0^2+m_1^2\bigr)\! + 24(m_0+m1) - 2\bigr)\!+ \cdots ,\cr
&\zeta_2(s, p; m_0, m_1)=\frac{s+1}{180 p^3} \bigl(1+10 \bigl(m_0^3+m_1^3+6(m_0+m_1)^2-(m0+m1)\bigr) \bigr) \cr
&\hphantom{\zeta_2(s, p; m_0, m_1)=}{}                \!+\frac{(s+1)(32s+105)}{9216 p^4}\bigl(m_0^2-m_1^2\bigr)\bigl(5\bigl(m_0^2+m_1^2\bigr)\!+24(m_0+m_1)-2\bigr)\!+\cdots.
\end{align*}

One can also use the formula~\eqref{1_Psi_Proposition} without expanding it around $\nu=\frac{1}{2}+\epsilon$. Instead, the function $\Psi_\nu(s, p)$ can be computed recursively order by order in $p$ to obtain the same expansion as in~\eqref{1_Psi_Expansion}.

\section[Asymptotic of omega for kappa=kappa']{Asymptotic of $\boldsymbol{\omega}$ for $\boldsymbol{\kappa=\kappa'}$} \label{section:Omega}
In this section, we are discussing the function $\omega^{{\rm sc}}$, appearing in the JMS theorem. Similar to the previous section, we first revisit the results of~\cite{Boos_TBAE, HGS_4, HGS_1, HGS_2, HGS_3} and then present an alternative expression for a function $\Theta(s, s'; p, \alpha)$, related to the function $\omega^{{\rm sc}}$.

We restrict our consideration to the case $\kappa=\kappa'$, so that $\rho^{{\rm sc}}(\lambda; \kappa, \kappa')=1$. Our goal is to give an algorithm for deriving the asymptotic expansion of the function $\omega^{{\rm sc}}(\lambda, \mu; \kappa, \kappa, \alpha)$. It satisfies the integral representation~\cite{BG_2009}
\begin{align}\label{2_omega}
\omega^{{\rm sc}}(\lambda, \mu; \kappa, \kappa, \alpha)=(f_{{\rm left}}\star f_{{\rm right}}+f_{{\rm left}}\star R_{{\rm dress}}\star f_{{\rm right}})(\lambda, \mu)+\omega_0(\lambda, \mu; \alpha),
\end{align}
with the functions defined as
\begin{align*}
&f_{{\rm left}}=\frac{1}{2\pi {\rm i}}\delta_\lambda^{-}\psi_0\biggl(\frac{\lambda}{\mu}, \alpha\biggr),\qquad
f_{{\rm right}}=                \delta_\mu^{-}    \psi_0\biggl(\frac{\lambda}{\mu}, \alpha\biggr),\\
&\omega_0(\lambda, \mu; \alpha)=  \delta_\lambda^{-}\delta_\mu^{-}\Delta_\lambda^{-1}\psi_0\biggl(\frac{\lambda}{\mu}, \alpha\biggr), \qquad
\psi_0(\lambda, \alpha)=\frac{\lambda^\alpha}{\lambda^2-1},
\end{align*}
and the operators defined as
\begin{align*}
\delta_\lambda^{-}f(\lambda)=f(q\lambda)-f(\lambda),\qquad
\Delta_\lambda f(\lambda)=f(q\lambda)-f\bigl(q^{-1}\lambda\bigr).
\end{align*}
The inverse of $\Delta_\lambda$ is understood as the principal value of the integral
\begin{align*}
\Delta^{-1}_\lambda \psi_0(\lambda, \alpha)=-\text{VP} \int_0^{\infty} \frac{\psi_0(\mu, \alpha)}{2\nu\bigl(1+(\lambda/\mu)^\frac{1}{\nu}\bigr)} \frac{{\rm d}\mu^2}{2\pi {\rm i}\mu^2}
\end{align*}
taken for the pole $\mu^2=1$. Finally, the contraction $\star$ denotes the integral
\begin{align*}
f \star g = \int_{\tilde{\gamma}(0, k)} f(\lambda) g(\lambda)\, {\rm d}m(\lambda)
\end{align*}
with the measure defined as
\begin{align*}
{\rm d}m(\lambda)=\frac{{\rm d}\lambda}{\lambda^2 \rho^{{\rm sc}} (\lambda; \kappa, \kappa) (1+\mathfrak{a}^{{\rm sc}}(\lambda, \kappa))}
           =\frac{{\rm d}\lambda}{\lambda^2 (1+\mathfrak{a}^{{\rm sc}}(\lambda, \kappa))}.
\end{align*}
The contour $\tilde{\gamma}(0, k)$ corresponds to the contour $\gamma(0, k)$ taken for the variable $\lambda^2$ instead of ${\zeta^2=\lambda^2 \bar{a}^{2\nu}}$. We remind the reader that the contour $\gamma(0, k)$ itself is taken around all Bethe roots (including the moved ones) in the clockwise direction.

The ``dressed'' resolvent appearing in~\eqref{2_omega} satisfies the integral equation
\begin{align*}%\label{2_Rd_IntEq}
R_{{\rm dress}}-R_{{\rm dress}}\star K_{0}=K_{0}, \qquad K_{0}(\lambda, \alpha)=\frac{1}{2\pi {\rm i}} \Delta_\lambda \psi_0(\lambda, \alpha).
\end{align*}
The solution of this integral equation was found in~\cite{HGS_4} in the vacuum case and generalised for the excited state in~\cite{Boos_TBAE}. Applying the same trick which was used to derive~\eqref{1_BAE_Int_Deform}, namely, deforming the integration contour and taking into account additional terms coming from the residues corresponding to particles and holes, one can obtain the expression for $R_{{\rm dress}}$ taking the ansatz
\begin{align}
R_{{\rm dress}} &= K_0(t/u, \alpha) \label{2_RDrees_Ansatz}\\
&\quad{}+\int_{-\infty}^{+\infty}\frac{{\rm d}l}{2\pi} \int_{-\infty}^{+\infty}\frac{{\rm d}m}{2\pi}
 t^{{\rm i}l} u^{{\rm i}m} S(l, \alpha) S(m, 2-\alpha) \hat{K}(l, \alpha) \hat{K}(m, 2-\alpha) \Theta(l, m; \kappa, \alpha).\nonumber
\end{align}
Here $\hat{K}(k, \alpha)$ is the image of the kernel $K_0(t, \alpha)$ under the Mellin transform
\begin{align*}
\hat{K}(k, \alpha)=\frac{\sinh\pi\bigl((2\nu-1)k-\frac{{\rm i}\alpha}{2}\bigr)}{\sinh\pi\bigl(k+\frac{{\rm i}\alpha}{2}\bigr)},
\end{align*}
and the function $S(k, \alpha)$ is
\begin{align*}
S(k, \alpha)=\frac{\Gamma\bigl(1+(1-\nu){\rm i}k-\frac{\alpha}{2}\bigr) \Gamma\bigl(\frac{1}{2}+{\rm i}\nu k\bigr)}{\Gamma\bigl(1+{\rm i}k-\frac{\alpha}{2}\bigr) \sqrt{2\pi} (1-\nu)^{\sfrac{(1-\alpha)}{2}}} {\rm e}^{{\rm i}\delta k}.
\end{align*}
Together they  satisfy the factorisation property
\begin{align*}
1-\hat{K}(k, \alpha)=S(k, \alpha)^{-1} S(-k, 2-\alpha)^{-1},
\end{align*}
where we assume that $0<\alpha<2$. Finally, the function $\Theta(l, m; \kappa, \alpha)$ has the following asymptotic expansion at $\kappa\rightarrow\infty$:
\begin{align*}%\label{2_Theta_Asymptotics}
\Theta(l, m; \kappa, \alpha) = \sum_{n=0}^{\infty} \Theta_n(l, m; \alpha) \kappa^{-n},
\qquad
\Theta_0(l, m; \alpha)=-\frac{{\rm i}}{l+m},
\end{align*}
and each mode $\Theta_n$ of this expansion can be calculated by iterations
\begin{gather}\label{2_Theta_Int}
\Theta(l, m; \kappa, \alpha)-\Theta_0(l, m; \alpha)=\frac{1}{f\kappa}\res_{l'}\res_{m'}
\Biggl[
S\bigl(l', 2-\alpha\bigr) S\bigl(m', \alpha\bigr) \frac{\Theta\bigl(m', m; \kappa, \alpha\bigr)}{l+l'}\cr
\qquad{}\times\Biggl(
-{\rm i}\sum_{r=1}^{k} \frac{{\rm e}^{\sfrac{-(l'+m')x_r^+(\kappa)}{f\kappa}}}{\bar{F}'\bigl(x_r^+(\kappa), \kappa\bigr)-1}
+{\rm i}\sum_{r=1}^{k} \frac{{\rm e}^{\sfrac{-(l'+m')x_r^-(\kappa)}{f\kappa}}}{\bar{F}'(x_r^-(\kappa), \kappa)-1} \cr
\hphantom{\times\Biggl(}\qquad{} +2\sum_{n=0}^{\infty}\frac{1}{n!} \int_{0}^{\infty} {\rm d}x
\biggl\{
{\rm e}^{-\sfrac{(l'+m')x}{f\kappa}} \bar{F}(x, \kappa)^n \biggl(-\frac{\partial}{\partial x}\biggr)^n
\biggr\}_{\text{odd}}
\frac{1}{1+{\rm e}^{2\pi x}}
\Biggr)
\Biggr].
\end{gather}
Here $\bar{F}'(x, \kappa)=\frac{\partial}{\partial x}\bar{F}(x, \kappa)$.

Performing iterations implies that a sufficient number of terms in the expansion of $\bar{F}(x, \kappa)$ and $x^{\pm}(\kappa)$ with respect to $\kappa^{-1}$ have been obtained. This can be done using~\eqref{1_FBar} and~\eqref{1_BetheRoots_Final}. For a few leading terms in the expansion for $\Theta(l, m; \kappa, \alpha)$, one can get
\begin{align*}
&\Theta(l, m; \kappa, \alpha)=\frac{-{\rm i}}{l+m}  \\
&\hphantom{\Theta(l, m; \kappa, \alpha)=}{}
+\frac{1}{(f\kappa)^2}\Biggl( \frac{1}{24\nu}-\frac{1}{2\nu}\sum_{r=1}^{k}\bigl(I^{(+)}_r+I^{(-)}_r\bigr)\Biggr)\!\biggl(-{\rm i}\nu(l+m)-\frac{1}{2}+\Delta_\alpha\biggr)+O\bigl(\kappa^{-3}\bigr), \nonumber
\end{align*}
where $\Delta_\alpha$ is given by~\eqref{Reminder_CFTDim}.

The connection between $\Theta$ and $\omega^{{\rm sc}}$ can be found with the help of the integral representation~\eqref{2_omega} and ansatz~\eqref{2_RDrees_Ansatz},
\begin{align*}
&\omega^{{\rm sc}}(\lambda, \mu; \kappa, \kappa, \alpha) \\
&\qquad{}\simeq\frac{1}{2\pi {\rm i}}\! \int_{-\infty}^{\infty} \! {\rm d}l \int_{-\infty}^{\infty}\! {\rm d}m
\tilde{S}(l, \alpha) \tilde{S}(m, 2-\alpha) \Theta(l+{\rm i}0, m; \kappa, \alpha)
\biggl(\frac{{\rm e}^{\delta+\pi {\rm i} \nu}\lambda^2}{\kappa^{2\nu} c(\nu)}\biggr)^{{\rm i}l}
\biggl(\frac{{\rm e}^{\delta+\pi {\rm i} \nu}\mu^2}{\kappa^{2\nu}  c(\nu)}\biggr)^{{\rm i}m}, \nonumber
\end{align*}
Here we returned to the variables $\lambda$ and $\mu$, and the function $\tilde{S}(k, \alpha)$ is given by
\begin{align*}
\tilde{S}(k, \alpha)=\frac{\Gamma\bigl(-{\rm i}k+\frac{\alpha}{2}\bigr) \Gamma\bigl(\frac{1}{2}+{\rm i}\nu k\bigr)}{\Gamma\bigl(-{\rm i}(1-\nu)k+\frac{\alpha}{2}\bigr) \sqrt{2\pi}(1-\nu)^{\sfrac{(1-\alpha)}{2}}}.
\end{align*}
The asymptotic expansion at $\lambda, \mu \rightarrow \infty$ can be obtained by computing the residues of the functions $\tilde{S}(l, \alpha)$ and $\tilde{S}(m, 2-\alpha)$,
\begin{align*}
\omega^{{\rm sc}}(\lambda, \mu; \kappa, \kappa, \alpha) \simeq
\sum_{r, s=1}^{\infty} \frac{1}{r+s-1} D_{2r-1}(\alpha) D_{2s-1}(2-\alpha)
\lambda^{-\frac{2r-1}{\nu}} \mu^{-\frac{2s-1}{\nu}}
\Omega_{2r-1, 2s-1}(p, \alpha),
\end{align*}
where
\begin{align*}
&D_{2n-1}(\alpha)=\frac{1}{\sqrt{{\rm i}\nu}} \Gamma(\nu)^{-\frac{2n-1}{\nu}} (1-\nu)^{\frac{2n-1}{2}} \frac{1}{(n-1)!}
\frac{\Gamma\bigl(\frac{\alpha}{2}+\frac{1}{2\nu}(2n-1)\bigr)}{\Gamma\bigl(\frac{\alpha}{2}+\frac{1-\nu}{2\nu}(2n-1)\bigr)},
\\
&\Omega_{2r-1, 2s-1}(p, \alpha)=
-\Theta\biggl(\frac{{\rm i}(2r-1)}{2\nu}, \frac{{\rm i}(2s-1)}{2\nu}; p, \alpha \biggr)
\biggl(\frac{r+s-1}{\nu}\biggr)
\biggl(\frac{\sqrt{2}p\nu}{R}\biggr)^{2r+2s-2}.
\end{align*}

\subsection[Alternative form for the function Theta]{Alternative form for the function $\boldsymbol{\Theta}$}
The iterative procedure provided by~\eqref{2_Theta_Int} is cumbersome for the same reason~\eqref{1_Psi_1} is: it requires calculating the expansions of $\bar{F}(x, \kappa)$ and $x^{\pm}(\kappa)$. Similarly to the previous case, we want to give an alternative expression for the function $\Theta(l, m; \kappa, \alpha)$ that would depend on the coordinates of particle-hole excitations in a direct way. Once again, we will be working with ``renormalised'' variables and functions
\begin{align*}
\Theta_\nu\bigl(s, s'; p, \alpha\bigr):=\frac{1}{2\nu} \Theta\biggl(\frac{s}{2\nu {\rm i}}, \frac{s'}{2\nu {\rm i}}; p, \alpha\biggr),
\qquad
 S_\nu(h, \alpha):=S\biggl(\frac{h}{2\nu {\rm i}}, \alpha\biggr).
\end{align*}
We also introduce the Taylor series at $\nu=\frac{1}{2}+\epsilon$ with respect to $\epsilon$ for the function $\Theta(s, s'; p, \alpha)$
\begin{align*}
\Theta_\nu \bigl(s, s'; p, \alpha\bigr)=\sum_{i=0}^{\infty} \eta_i\bigl(s, s'; p, \alpha\bigr) \epsilon^i,
\end{align*}
and for the function $S(h, \alpha)$
\begin{align*}
S_\nu(h, \alpha)=1+\sum_{i=0}^{\infty} \sigma_i(h, \alpha) \epsilon^i.
\end{align*}
Note that unlike in~\eqref{Sh_Expansion_Epsilon} the zeroth term in the expansion of $S(h, \alpha)$ with respect to $\epsilon$ is not just 1 but rather
\begin{align*}
1+\sigma_0(h, \alpha)=\frac{\Gamma\bigl(1+\frac{h}{2}-\frac{\alpha}{2}\bigr) \Gamma\bigl(\frac{1}{2}+\frac{h}{2}\bigr)}{\Gamma\bigl(\frac{1}{2}+\frac{h}{2}-\frac{\alpha}{4}\bigr) \Gamma\bigl(1+\frac{h}{2}-\frac{\alpha}{4}\bigr)}=
1+\frac{\alpha(\alpha-2)}{8h}+\cdots.
\end{align*}
Now we are ready to make another proposition.

\begin{Proposition}
\label{prop:2}
The function $\Theta_\nu(s, s'; p, \alpha)$ is given by the formula
\begin{align}
0&{}=\res_h \res_{h'} \frac{S_{\nu}(h, 2-\alpha) S_\nu\bigl(h', \alpha\bigr)}{s+h} \Theta_\nu\bigl(h', s'; p, \alpha\bigr)\nonumber \\
&\quad{}\times \langle 0| \exp\bigl\{\res_{h''}\bigl[\bigl(S_\nu\bigl(h''\bigr)-1\bigr)\Psi_\nu\bigl(h'', p\bigr)\Phi_{+}\bigl(h''\bigr)\bigr] \bigr\} \mathcal{G}\bigl(-h-h', p, \Phi_-\bigr) |0\rangle, \label{Theta_Proposition}
\end{align}
where $\mathcal{G}(s, p, \Phi_-)$ is a functional of fields $\Phi_{-}$,
\begin{align*}
&\mathcal{G}(s, p, \Phi_-)=
 \sum_{n=0}^{\infty}\frac{p^{-n}}{n!}
                    \res_{h_1}\dots\res_{h_n} \Phi_-(h_1)\dots\Phi_-(h_n)
                    \Bigl(1+s-\sum h_j\Bigr)_{n+1} \\
&\hphantom{\mathcal{G}(s, p, \Phi_-)=\sum_{n=0}^{\infty}}{}
\times\zeta_0\Bigl(n+1+s-\sum h_j, p\Bigr).
\end{align*}
\end{Proposition}

The proof of Proposition \ref{prop:2} is similar to the proof of Proposition \ref{prop:Psi} given in Appendix~\ref{appendix:PsiProof}. Here we only make some comments. The formula~\eqref{Theta_Proposition} can be brought to the form similar to~\eqref{1_Psi_Proposition} by replacing the function $\zeta_0(1+s-\sum h_j, p)$ in the expression for $\mathcal{G}(s, p, \Phi_-)$ with the regular one $\zeta_R(1+s-\sum h_j, p)$,
\begin{align*}
&\Theta_\nu(s, s'; p, \alpha) - \frac{1}{s+s'} = \res_h \res_{h'} \frac{S_{\nu}(h, 2-\alpha) S_\nu\bigl(h', \alpha\bigr)}{s+h} \Theta_\nu\bigl(h', s'; p, \alpha\bigr) \\
&\qquad{}\times\langle 0| \exp\bigl\{\res_{h''}\bigl[\bigl(S_\nu\bigl(h''\bigr)-1\bigr)\Psi_\nu\bigl(h'', p\bigr)\Phi_{+}\bigl(h''\bigr)\bigr] \bigr\} \mathcal{G}_R\bigl(-h-h', p, \Phi_-\bigr) |0\rangle \nonumber,
\end{align*}
with
\begin{align*}
\begin{split}
&\mathcal{G}_R(s, p, \Phi_-)=
 \sum_{n=0}^{\infty}\frac{p^{-n}}{n!}
                    \res_{h_1}\cdots\res_{h_n} \Phi_-(h_1)\cdots\Phi_-(h_n)
                    \Bigl(1+s-\sum h_j\Bigr)_{n+1} \\
&\hphantom{\mathcal{G}_R(s, p, \Phi_-)=\sum_{n=0}^{\infty}}{}
\times\zeta_R\Bigl(n+1+s-\sum h_j, p\Bigr).
\end{split}
\end{align*}
In this way the proof of Proposition \ref{prop:2} is even more straightforward since one does not have to obtain the singular term as it was done for the function $\Psi(s, p)$ in Appendix~\ref{appendix:PsiProof}.

From~\eqref{Theta_Proposition} follow the equations for the functions $\eta_i(s, s'; p, \alpha)$,
\begin{gather*}
 \eta_0\bigl(s, s'; p, \alpha\bigr)\\
\qquad{}=\res_h \res_{h'} \frac{\sigma_0(h, 2-\alpha) \sigma_0\bigl(h', \alpha\bigr)}{s+h} \eta_0\bigl(h', s'; p, \alpha\bigr) p^{-1} \bigl(1-h-h'\bigr) \zeta_0\bigl(1-h-h', p\bigr), \\
 \eta_1\bigl(s, s'; p, \alpha\bigr)\\
 \qquad{}=\res_h \res_{h'} \frac{\sigma_1(h, 2-\alpha) \sigma_0\bigl(h', \alpha\bigr)}{s+h} \eta_0\bigl(h', s'; p, \alpha\bigr) p^{-1} \bigl(1-h-h'\bigr) \zeta_0\bigl(1-h-h', p\bigr) \\
 \qquad\hphantom{=}{}+\res_h \res_{h'} \frac{\sigma_0(h, 2-\alpha) \sigma_1\bigl(h', \alpha\bigr)}{s+h} \eta_0\bigl(h', s'; p, \alpha\bigr) p^{-1} \bigl(1-h-h'\bigr) \zeta_0\bigl(1-h-h', p\bigr) \\
 \qquad\hphantom{=}{}+\res_h \res_{h'} \frac{\sigma_0(h, 2-\alpha) \sigma_0\bigl(h', \alpha\bigr)}{s+h} \eta_1\bigl(h', s'; p, \alpha\bigr) p^{-1} \bigl(1-h-h'\bigr) \zeta_0\bigl(1-h-h', p\bigr)\\
 \qquad\hphantom{=}{}+\res_h \res_{h'} \biggl(\frac{\sigma_0(h, 2-\alpha) \sigma_0\bigl(h', \alpha\bigr)}{s+h} \eta_0\bigl(h', s'; p, \alpha\bigr) \\
 \qquad\hphantom{=+\res_h \res_{h'} }{}\times\res_{h''} \bigl(\sigma_1\bigl(h''\bigr) \zeta_0\bigl(h'', p\bigr) p^{-2} \bigl(1-h-h'-h''\bigr)_2 \zeta_0\bigl(2-h-h'-h'', p\bigr) \bigr) \biggl).
\end{gather*}
They can be easily computed in each order with respect to variable $p$. For example, for a single particle-hole excitation $I^{(+)}=\{m_0\}$, $I^{(-)}=\{m_1\}$ we can calculate
\begin{gather*}
\eta_0\bigl(s, s'; p, \alpha\bigr) =\frac{1}{s+s'} + \frac{1}{96 p^2} (-1+12m_0+12m_1) \bigl(4\bigl(1+s+s')+\alpha(2-\alpha)\bigr) \\
\hphantom{\eta_0(s, s'; p, \alpha) =}{}
+\frac{1}{512p^3} \bigl(m_0^2-m_1^2\bigr) \bigl(64\bigl(1+s+s'\bigr)\bigl(2+s+s'\bigr)+48\bigl(2+s+s'\bigr)\alpha \cr
\hphantom{\eta_0(s, s'; p, \alpha) =}{}
-12\bigl(3+2s+2s'\bigr)\alpha^2-12\alpha^3+3\alpha^4\bigr)+\cdots ,\cr
\eta_1\bigl(s, s'; p, \alpha\bigr) =\frac{1}{16 p^2}(-1+12m_0+12m_1)\alpha(2-\alpha) \cr
\hphantom{\eta_1(s, s'; p, \alpha) =}{}                         +\frac{9}{128 p^3}\bigl(m_0^2-m_1^2\bigr)\alpha(2-\alpha)\bigl(4\bigl(2+s+s'\bigr)+\alpha(2-\alpha)\bigr)+\cdots . \nonumber
\end{gather*}

\begin{Remark}
Consider the expression~\eqref{Theta_Proposition} in the free fermionic case. The exponential ``dressing'' disappears in the case $\nu=1/2$ leaving us with the following integral equation for the function $\Theta_{1/2}(s, s'; p, \alpha)$,
\begin{align}\label{2_Theta_FF}
\Theta_{1/2}\bigl(s, s'; p, \alpha\bigr)-\frac{1}{s+s'}=\res_h \mathcal{K}(s, h; p, \alpha) S_{1/2}(h, \alpha) \Theta_{1/2}\bigl(h, s'; p, \alpha\bigr)
\end{align}
with the kernel
\begin{align*}
\mathcal{K}(s, s'; p, \alpha)=p^{-1}\res_h \frac{S_{1/2}(h, 2-\alpha) \bigl(1-h-s'\bigr) \zeta_R\bigl(1-h-s', p\bigr)}{s+h}.
\end{align*}
At the same time, in the free fermionic case there exists an explicit representation for the function~$\omega^{{\rm sc}}(\lambda, \mu; \kappa, \kappa', \alpha)$ \cite{Smirnov_PC}. We believe it is possible to relate this result to the integral equation~\eqref{2_Theta_FF}, which, hopefully, will help us to generalise the above formulas for $\omega(\lambda, \mu; \kappa, \alpha)$ for the case of $\kappa\neq\kappa'$.
\end{Remark}

\section{Conclusion and discussions}
In this paper, we introduced a new representation for the functions $\Psi(l, \kappa)$, related to the solutions of the Bethe ansatz equations, and $\Theta(l, m; \kappa, \alpha)$, related to the function $\omega^{{\rm sc}}$. The key advantage of these representations lies in the explicit dependence of the functions on the coordinates of particle-hole excitations. This feature reduces the number of calculations required to determine these functions in the iterative procedure, progressing order by order in the variable~$p$. We hope that these formulae will be useful to study the case of different left and right states $\kappa\neq\kappa'$. Form factors correspond to a special case when one of the states is a vacuum state. Since for form factors many results are already known, we believe it would be interesting to generalise the above approach to the case of form factors as well.

\appendix
\section{Proof of Proposition \ref{prop:Psi}}\label{appendix:PsiProof}
In this appendix, we give the proof of Proposition \ref{prop:Psi}. For simplicity, we consider the case without any particle-hole excitations. We want to show that the equation~\eqref{1_Psi_Proposition} gives the same result as the equation~\eqref{1_Psi_1} in every order of the $\epsilon$-expansion. In ``renormalised'' variables~\eqref{1_Rescale}, the representation~\eqref{1_Psi_1} of $\Psi_\nu$ takes the form
\begin{align}
&\Psi_\nu(s, p)-p\Psi_0(s, p) \nonumber\\
&\qquad{}=-\frac{2}{p} \sum_{n=0}^{\infty} \frac{1}{n!} \int_{0}^{\infty} \frac{{\rm d}x}{2\pi}
        \biggl\{\res_h\biggl[ \frac{{\rm e}^{\frac{{\rm i}hx}{p}}}{s+h} S(h) \biggr] \bar{F}(x, p)^n \biggl(-\frac{\partial}{\partial x}\biggr)^n \biggr\}_{\text{even}}\log\bigl(1+{\rm e}^{-2\pi x}\bigr),\label{A_Psi_HGS}
\end{align}
where the function $\bar{F}(x, p)$ is given by
\begin{align}\label{A_1}
\bar{F}(x, p)=x+{\rm i}\res_h\big[ {\rm e}^{\frac{{\rm i}hx}{p}} S(h)\Psi_\nu(h, p) \big].
\end{align}
Let us introduce the expansion of the function $\bar{F}(x, p)$ with respect to $\epsilon$,
\begin{align*}
\bar{F}(x, p)=\sum_{j=0}^{\infty} \bar{F}_j(x, p) \epsilon^j.
\end{align*}
From~\eqref{A_1}, it follows that
\begin{align}
&\bar{F}_0(x, p)=x+{\rm i}\res_h\bigl[ {\rm e}^{\frac{{\rm i}hx}{p}} \zeta_0(h, p) \bigr] = x+\res_h\biggl[ {\rm i} {\rm e}^{\frac{{\rm i}hx}{p}} \frac{p}{h(h-1)} \biggr] = {\rm i}p \sum_{m=2}^{\infty} \frac{1}{m!} \biggl(\frac{{\rm i}x}{p}\biggr)^m, \nonumber\\
&\bar{F}_1(x, p)={\rm i}\res_h\bigl[ {\rm e}^{\frac{{\rm i}hx}{p}} \sigma_1(h)\zeta_0(h, p) \bigr], \cr
&\bar{F}_2(x, p)={\rm i}\res_h\bigl[ {\rm e}^{\frac{{\rm i}hx}{p}} (\sigma_2(h)\zeta_0(h, p) + \sigma_1(h)\zeta_1(h, p)) \bigr], \qquad \dots.
\label{A_F_Epsilon}
\end{align}
The $\epsilon$-expansion of the $\Psi(h, p)$ can be obtained from~\eqref{A_Psi_HGS},
\begin{gather}
\Psi_\nu(s, p)-p\Psi_0(s, p)\nonumber\\
\qquad{}=-\frac{2}{p} \sum_{n=0}^{\infty} \frac{1}{n!} \int_{0}^{\infty} \frac{{\rm d}x}{2\pi}
        \bigg\{ \res_h\biggl[ \frac{{\rm e}^{\frac{{\rm i}hx}{p}}}{s+h} \biggr]
        \bigg(F_0(x, p)^n
              \epsilon\cdot nF_0(x, p)^{n-1}F_1(x, p)  \cr
\qquad\hphantom{=}{} +\epsilon^2 \biggl(nF_0(x, p)^{n-1}F_2(x, p) + \frac{n(n-1)}{2}F_0(x, p)^{n-1}F_1(x, p)^2\biggr) + \cdots \bigg) \! \biggl(-\frac{\partial}{\partial x}\biggr)^n \bigg\}_{\text{even}} \cr
\qquad\hphantom{=+}{}\times \log\bigl(1+{\rm e}^{-2\pi x}\bigr).\label{A_HGS_Epsilon}
\end{gather}

As it can be seen, in every order of the $\epsilon$-expansion of the function $\Psi(s, p)$ one has to calculate the integral of the form
\begin{align}\label{A_Int}
-\frac{2}{p} \sum_{n=0}^{\infty} \frac{1}{n!} \int_{0}^{\infty} \frac{{\rm d}x}{2\pi} \biggl\{ {\rm e}^{-\frac{{\rm i}sx}{p}} \bar{F}_0(x, p)^n \biggl(-\frac{\partial}{\partial x}\biggr)^{n+n'} \biggr\}_{\text{even}} \log\bigl(1+{\rm e}^{-2\pi x}\bigr)
\end{align}
for some $n'\geq 0$. To do so, let us introduce the following notation:
\begin{align*}%\label{A_u}
\frac{1}{n!}\Biggl[ \sum_{k=2}^{\infty} \frac{x^k}{k!} \Biggr]^n=
\sum_{m=2n}^{\infty} u_{n, m} x^m.
\end{align*}
The numbers $u_{n, m}$ can be written down explicitly in the form
\begin{align*}
u_{n, m}=\frac{1}{n!} \sum_{\substack{2\leq k_1 \leq\cdots\leq k_n \leq m-2(n-1), \\ k_1+\cdots+k_n=m}} \frac{1}{k_1! \cdots k_n!}.
\end{align*}
The most important property of $u_{n, m}$ relevant to this proof is the following sum:
\begin{align}\label{A_u_1}
\sum_{\substack{0 \leq n \leq r, \\ 0 \leq k \leq n+r}} (-1)^{n+k} \frac{(n+r)!}{k!} u_{n, r+n-k} s^k=(-1)^r (s+1)_r.
\end{align}
Then the left-hand side of the formula~\eqref{A_Int} can be written as
\begin{align}
&-\frac{2}{p} \sum_{n=0}^{\infty} \frac{1}{n!} \int_{0}^{\infty} \frac{{\rm d}x}{2\pi} \biggl\{ {\rm e}^{-\frac{{\rm i}sx}{p}} \bar{F}_0(x, p)^n \biggl(-\frac{\partial}{\partial x}\biggr)^{n+n'} \biggr\}_{\text{even}} \log\bigl(1+{\rm e}^{-2\pi x}\bigr)\nonumber\\
&\qquad{}=-\frac{2}{p} \sum_{\substack{n \geq 0, \\ k \geq 0, \\ m \geq 2n}} \int_{0}^{\infty} \frac{{\rm d}x}{2\pi}
    \biggl\{{\rm i}^{-k+n+m} p^{n-m-k} x^{k+m} \frac{s^k}{k!} u_{n, m} \biggl(-\frac{\partial}{\partial x}\biggr)^{n+n'} \biggr\}_{\text{even}} \log\bigl(1+{\rm e}^{-2\pi x}\bigr)\cr
&\qquad{} =-2\sum_{\substack{n \geq 0,\\ k \geq 0, \\ m \geq 2n}} {\rm i}^{-k+n+m} p^{n-m-k-1} \frac{s^k}{k!} u_{n, m} (m+k)! \bigl(1-2^{-m-k-1+n+n'}\bigr)\nonumber\\
&\qquad\hphantom{=-2\sum_{\substack{n \geq 0,\\ k \geq 0, \\ m \geq 2n}}}{}
\times\frac{\zeta\bigl(m+k-n-n'+2\bigr)}{(2\pi)^{m+k-n-n'+2}}.\label{A_2}
\end{align}
In the final line, the equality
\begin{align*}
\int_{0}^{\infty} \frac{{\rm d}x}{2\pi} x^{m} \biggl(-\frac{\partial}{\partial x}\biggr)^n \log\bigl(1+{\rm e}^{-2\pi x}\bigr) = m!\bigl(1-2^{-m-1+n}\bigr) \frac{\zeta(m-n+2)}{(2\pi)^{m-n+2}}
\end{align*}
was used and it was assumed that $k+m-n-n'$ is even.

Let us denote $m+k-n=r$. Since $u_{n, m}=0$ for $m<2n$, the sums over $k$ and $n$ in~\eqref{A_2} are actually finite. Using the property~\eqref{A_u_1} of the numbers $u_{n, m}$, we continue the derivation
\begin{align}
\eqref{A_2}&{}=-2\sum_{\substack{n \geq 0,\\ k \geq 0,\\ r \geq 0}} p^{-r-1} (-1)^{n+k} {\rm i}^{r} \frac{(r+n)!}{k!} u_{n, r-k+n} s^k \bigl(1-2^{-r+n'-1}\bigr) \frac{\zeta\bigl(r-n'+2\bigr)}{(2\pi)^{r-n'+2}}\nonumber\\
           &{}=-2\sum_{\substack{0 \leq n \leq r,\\ 0 \leq k \leq r-n, \\ r \geq 0}} p^{-r-1} (-1)^{n+k} {\rm i}^{r} \frac{(r+n)!}{k!} u_{n, r-k+n} s^k \bigl(1-2^{-r+n'-1}\bigr) \frac{\zeta\bigl(r-n'+2\bigr)}{(2\pi)^{r-n'+2}}\cr
           &{}=-2\sum_{r \geq 0} (-{\rm i})^{r} p^{-r-1} (s+1)_r \bigl(1-2^{-r-n'+1}\bigr)\frac{\zeta\bigl(r-n'+2\bigr)}{(2\pi)^{r-n'+2}}.\label{A_3}
\end{align}
We can denote $r-n'+2=2m$ since it is even. Then
\begin{align*}
\eqref{A_3}&=-2\sum_{m=-n'+2}^{\infty} (-{\rm i})^{2m+n'-2} p^{-2m+1-n'} (s+1)_{2m+n'-2} \bigl(1-2^{-2m+1}\bigr)\frac{\zeta(2m)}{(2\pi)^{2m}}\\
                     &=-2({\rm i}p)^{-n'}(1+s)_{n'}\frac{1}{s+n'}\sum_{m=1}^{\infty} p^{-2m+1} \bigl(s+n'\bigr)_{2m-1} \bigl(1-2^{-2m+1}\bigr) (-{\rm i})^{2m-2} \frac{|B_{2m}|}{2}\cr
                     &\quad{} -2({\rm i}p)^{-n'} \sum_{m=-n'+2}^{0} {\rm i}^{2m-2} p^{-2m+1} (1+s)_{2m+n'-2} \bigl(1-2^{-2m+1}\bigr)\frac{\zeta(2m)}{(2\pi)^{2m}} \cr
                     &=({\rm i}p)^{-n'} (1+s)_{n'} \zeta_{R}\bigl(s+n', p\bigr)-{\rm i}({\rm i}p)^{-n'+1}(1+s)_{n'-2}\cr
                     &\quad{} -2({\rm i}p)^{-n'} \sum_{m=-n'+2}^{0} {\rm i}^{2m-2} p^{-2m+1} (1+s)_{2m+n'-2} \bigl(1-2^{-2m+1}\bigr)\frac{\zeta(2m)}{(2\pi)^{2m}} .\nonumber
\end{align*}
For $n'\geq2$, the only term that contributes to the sum in the last line is the one with $m=0$. Otherwise the sum is zero. Therefore, we get to the following result:
\begin{align}
&-\frac{2}{p} \sum_{n=0}^{\infty} \frac{1}{n!} \int_{0}^{\infty} \frac{{\rm d}x}{2\pi} \biggl\{ {\rm e}^{-\frac{{\rm i}sx}{p}} \bar{F}_0(x, p)^n \biggl(-\frac{\partial}{\partial x}\biggr)^{n+n'} \biggr\}_{\text{even}} \log\bigl(1+{\rm e}^{-2\pi x}\bigr) \cr
&\qquad =
\begin{cases}
({\rm i}p)^{-n'} (1+s)_{n'} \zeta_{R}\bigl(s+n', p\bigr) & \text{for } n'=0, 1, \\
({\rm i}p)^{-n'} (1+s)_{n'} \zeta_{0}\bigl(s+n', p\bigr) & \text{for } n'\geq 2.
\end{cases}\label{A_Integral_Answer}
\end{align}

From this result for $n'=0$, it follows immediately that the $\epsilon^0$-term of the expansion~\eqref{A_Psi_HGS} is~$\zeta_0(s, p)$ which is in agreement with Proposition \ref{prop:Psi}. Now let us turn our attention to the first order with respect to $\epsilon$. From~\eqref{A_HGS_Epsilon} and~\eqref{A_F_Epsilon}, we get
\begin{align*}
\zeta_1(s, p)
&=-\frac{2}{p}\Biggl( \sum_{n=1}^{\infty} \frac{{\rm i}}{n!} \int_{0}^{\infty} \frac{{\rm d}x}{2\pi}
\biggl\{
    {\rm e}^{-\frac{{\rm i}sx}{p}} n \bar{F}_0(x, p)^{n-1} \res_h\biggl[ {\rm e}^{\frac{{\rm i}hx}{p}} \sigma_1(h)\zeta_0(h, p) \biggr]
    \biggl(-\frac{\partial}{\partial x}\biggr)^{n}
\biggr\}_{\text{even}}\cr
&\quad{}+\sum_{n=0}^{\infty} \frac{1}{n!} \int_{0}^{\infty} \frac{{\rm d}x}{2\pi}
\biggl\{
    \res_h\biggl[\frac{{\rm e}^{\frac{{\rm i}hx}{p}}}{s+h}\sigma_1(h)\biggr] \bar{F}_0(x, p)^n
    \biggl(-\frac{\partial}{\partial x}\biggr)^{n}
\biggr\}_{\text{even}}\Biggr) \log\bigl(1+{\rm e}^{-2\pi x}\bigr).
\end{align*}
We can shift the summation index in the first sum and then use~\eqref{A_Integral_Answer} to calculate both sums. This gives us
\begin{align}\label{A_Zeta1_1}
\zeta_1(s, p)=\res_h\big[ \sigma_1(h)\zeta_0(h, p) p^{-1} (1+s-h) \zeta_R(1+s-h, p) \big] + \res_h\biggl[ \frac{\sigma_1(h)\zeta_R(-h)}{s+h} \biggr].
\end{align}
To bring the answer to the final form we can use the duality property of $S(h)$ which in the new variables simply reads
\begin{align}
S(h)S(-h)=1 \quad\Rightarrow\quad &\sigma_1(h)+\sigma_1(-h)=0, \nonumber\\
                             &\sigma_1(h)\sigma_1(-h)+\sigma_2(h)+\sigma_2(-h)=0, \cr
                             &\ldots.\label{A_S_Duality}
\end{align}
Then the last term in the formula~\eqref{A_Zeta1_1} can be modified and added to the first one,
\begin{align*}
\zeta_1(s, p)&{}=\res_h\bigl[ \sigma_1(h)\zeta_0(h, p) p^{-1} (1+s-h) \zeta_R(1+s-h, p) \bigr] + \res_h\biggl[ \frac{\sigma_1(h)\zeta_R(h)}{s-h} \biggr]  \\
&{}=\res_h\bigl[ \sigma_1(h)\zeta_0(h, p) p^{-1} (1+s-h) \zeta_R(1+s-h, p) \bigr] + \res_h\biggl[ \frac{\sigma_1(h)\zeta_0(h)}{s-h} \biggr]\\
&{}=\res_h\bigl[ \sigma_1(h)\zeta_0(h, p) p^{-1} (1+s-h) \zeta_0(1+s-h, p) \bigr].
\end{align*}
That is exactly the result~\eqref{1_Zetas} we obtain from the exponential representation.

To proceed with higher orders of the expansion of $\Psi_\nu(s, p)$, let us introduce a shorthand notation for the integral~\eqref{A_Psi_HGS} in form of the ``contraction''
\begin{align*}%\label{A_Contraction_Def}
&\langle g, f \rangle_{n'} \\
&\quad{}=-\frac{2}{p} \sum_{n=0}^{\infty} \frac{1}{n!} \int_{0}^{\infty} \frac{{\rm d}x}{2\pi}
    \biggl\{
        \res_h\biggl[{\rm e}^{\frac{{\rm i}hx}{p}} \frac{f(h)}{s+h} \bar{F}_0(x, p)^n f(x, p)\biggr]
        \biggl(-\frac{\partial}{\partial x}\biggr)^{n+n'}
    \biggr\}_{\text{even}} \!\log\bigl(1+{\rm e}^{-2\pi x}\bigr). \nonumber
\end{align*}
Then the formula~\eqref{A_Integral_Answer}, for example, takes a simple form
\begin{align*}
\langle 1, 1 \rangle_{n'} =
\begin{cases}
({\rm i}p)^{-n'} (1+s)_{n'} \zeta_{R}\bigl(s+n', p\bigr) & \text{for } n'=0, 1, \\
({\rm i}p)^{-n'} (1+s)_{n'} \zeta_{0}\bigl(s+n', p\bigr) & \text{for } n'\geq 2.
\end{cases}
\end{align*}
For $\zeta_1(s, p)$, we have
\begin{align*}
\zeta_1(s, p)=\bigl\langle 1, \bar{F}_1\bigr\rangle_1 + \langle \sigma_1, 1 \rangle_0.
\end{align*}
We apply~\eqref{A_S_Duality} to the last term and add to the first one to replace $\zeta_R(1+s, p)$ by $\zeta_0(1+s, p)$ as it was done above.

Now let us discuss the general case $\zeta_k(s, p)$. We will prove Proposition \ref{prop:Psi} by induction. Suppose it is correct for some $k-1$. Using the ``contraction'' defined above, we write
\begin{align}
\zeta_k(s, p)&= \bigl\langle 1, \bar{F}_{k} \bigr\rangle_1 + \bigl\langle 1, \bar{F}_{k-1} \bar{F}_1 \bigr\rangle_2 + \bigl\langle 1, \bar{F}_{k-2} \bar{F}_{2} \bigr\rangle_2 + \frac{1}{2!} \bigl\langle 1, \bar{F}_{k-2}\bar{F}_1^2\bigr\rangle + \cdots \nonumber\\
             &\quad{}+ \langle \sigma_k(h), 1 \rangle_0
              + \bigl\langle \sigma_{k-1}(h), \bar{F}_1 \bigr\rangle_1
              + \biggl( \bigl\langle \sigma_{k-2}, \bar{F}_2\bigr\rangle_1 + \frac{1}{2}\bigl\langle \sigma_{k-2}, \bar{F}_1^2 \bigr\rangle_2 \biggr)+ \cdots .\label{A_ZetaK_Gen}
\end{align}
The very first term is
\begin{align}
&\bigl\langle 1, \bar{F}_{k} \bigr\rangle_1 \nonumber\\
&\qquad{}= \res_h \bigl[ (\sigma_1(h)\zeta_{k-1}(h)+\cdots+\sigma_{k-1}(h)\zeta_1(h, p))(1+s-h) p^{-1} \zeta_R(1+s-h, p) \bigr].\label{A_1_FK}
\end{align}
If it had the function $\zeta_0(1+s-h, p)$ instead of $\zeta_R(1+s-h, p)$, the first line would have given us the exponent~\eqref{1_Psi_Proposition}, which is the direct consequence of the integral~\eqref{A_Integral_Answer}. It can be shown that the terms coming from the second line of~\eqref{A_ZetaK_Gen} are exactly the missing singular terms we need to add to~\eqref{A_1_FK} to replace $\zeta_R(1+s-h)$ by $\zeta_0(1+s-h, p)$. Indeed, let us apply the duality~\eqref{A_S_Duality},
\begin{align}
\langle \sigma_k, 1 \rangle_0 &{}= \res_h \biggl[ \frac{\sigma_k(h)}{s+h} \zeta_0(-h, p)\biggr] \nonumber\\
                                 &{}= \res_h \biggl[ \frac{\sigma_k(h)}{s-h} \zeta_0(h, p)\biggr] + \sum_{i=1}^{k-1} \res_h \biggl[ \frac{\sigma_i(h) \sigma_{k-i}(-h)}{s-h} \zeta_0(h, p)\biggr].\label{A_S_Duality_Gen}
\end{align}
Now in~\eqref{A_ZetaK_Gen} take the terms which are the ``contractions'' of $\sigma_i(h)$ with some powers of $\bar{F}_j$. Once again, we will get the function $\zeta_R$ in one of such ``contractions''
\begin{align*}
\begin{split}
&\bigl\langle \sigma_i, \bar{F}_{k-i} \bigr\rangle_1=\res_h\res_{h'} \biggl[ \frac{\sigma_i(h)}{s+h}\bigl(\sigma_{k-i}\bigl(h'\bigr)\zeta_0(h', p) + \cdots+ \sigma_1\bigl(h'\bigr)\zeta_{k-i-1}\bigl(h'\bigr)\bigr)\\
&\hphantom{\bigl\langle \sigma_i, \bar{F}_{k-i} \bigr\rangle_1=\res_h\res_{h'} \biggl[}{}
\times\bigl(1-h-h'\bigr)p^{-1} \zeta_R\bigl(1-h-h', p\bigr) \biggr].
\end{split}
\end{align*}
The missing singular term is exactly the term proportional to $\sigma_i(h)$ coming from~\eqref{A_S_Duality_Gen}. We write, therefore,
\begin{align*}
&\bigl\langle \sigma_i, \bar{F}_{k-i} \bigr\rangle_1 + \res_h \biggl[ \frac{\sigma_i(h) \sigma_{k-i}(-h)}{s-h} \zeta_0(h, p)\biggr] = \res_h\res_{h'} \biggl[ \frac{\sigma_i(h)}{s-h}\\
&\qquad{}\times\bigl(\sigma_{k-i}\bigl(h'\bigr)\zeta_0\bigl(h', p\bigr) + \cdots + \sigma_1\bigl(h'\bigr)\zeta_{k-i-1}\bigl(h'\bigr)\bigr)\bigl(1-h-h'\bigr)\frac{1}{p}\zeta_0\bigl(1-h-h', p\bigr) \biggr].
\end{align*}
Now, by the assumption of induction all terms the terms which are a ``contraction'' of $\sigma_i(h)$ with some powers of $\bar{F}_j$ sum up into
\begin{align*}
\sum_{i=1}^{k-1} \res_h\biggl[\frac{\sigma_i(h) \zeta_{k-i}(h, p)}{s-h} \biggr],
\end{align*}
and these are exactly the singular terms we need to be able to replace $\zeta_R$ by $\zeta_0$ in~\eqref{A_1_FK}. This observation finishes the proof of Proposition \ref{prop:Psi}.

\subsection*{Acknowledgements}
SA and HB acknowledge financial support from the DFG in the framework of the research unit FOR 2316 and through project BO 3401/8-1. The authors would like to thank F.~Smirnov for explaining his result for the case of different parameters $\kappa$ and $\kappa'$ at the free fermion point. The authors would also like to thank the anonymous referees for their valuable contributions in improving the paper. HB would also like to thank S.~Lukyanov for stimulating discussions.

\pdfbookmark[1]{References}{ref}
\LastPageEnding

\end{document}